\documentclass[pra,reprint,superscriptaddress,longbibliography]{revtex4-1}
\usepackage[utf8]{inputenc}
\usepackage[T1]{fontenc}
\usepackage{amssymb,amsthm,amsmath,bbm,bbold,graphicx,epsfig,threeparttable,color}
\usepackage{ulem,enumerate} 
\usepackage{enumitem}
\usepackage[title]{appendix}
\usepackage{hyperref}

\newcommand{\ket}[1]{{|#1\rangle}}
\newcommand{\bra}[1]{{\langle#1|}}

\newcommand{\Tr}{{\rm Tr}}

\theoremstyle{definition}

\begin{document}

\title{Unveiling the nonclassicality within quasi-distribution representations through deep learning}
\author{Hong-Bin Chen}
\affiliation{Department of Engineering Science, National Cheng Kung University, Tainan 701401, Taiwan}
\affiliation{Center for Quantum Frontiers of Research \& Technology, NCKU, Tainan 701401, Taiwan}
\affiliation{Physics Division, National Center for Theoretical Sciences, Taipei 106319, Taiwan}
\author{Cheng-Hua Liu}
\affiliation{Department of Engineering Science, National Cheng Kung University, Tainan 701401, Taiwan}
\author{Kuan-Lun Lai}
\affiliation{Department of Engineering Science, National Cheng Kung University, Tainan 701401, Taiwan}
\affiliation{Physics Division, National Center for Theoretical Sciences, Taipei 106319, Taiwan}
\author{Bor-Yann Tseng}
\affiliation{Department of Engineering Science, National Cheng Kung University, Tainan 701401, Taiwan}
\author{Ping-Yuan Lo}
\affiliation{Department of Electrophysics, National Yang Ming Chiao Tung University, Hsinchu 300093, Taiwan}
\author{Yueh-Nan Chen}
\affiliation{Department of Physics, National Cheng Kung University, Tainan 701401, Taiwan}
\affiliation{Center for Quantum Frontiers of Research \& Technology, NCKU, Tainan 701401, Taiwan}
\affiliation{Physics Division, National Center for Theoretical Sciences, Taipei 106319, Taiwan}
\author{Chi-Hua Yu}
\email{jonnyyu@gs.ncku.edu.tw}
\affiliation{Department of Engineering Science, National Cheng Kung University, Tainan 701401, Taiwan}

\begin{abstract}
To unequivocally distinguish genuine quantumness from classicality, a widely adopted approach focuses on the negative values of a quasi-distribution representation as compelling evidence of
nonclassicality. Prominent examples include the dynamical process nonclassicality characterized by the canonical Hamiltonian ensemble representation (CHER) and the nonclassicality of quantum states
characterized by the Wigner function. However, to construct a multivariate joint quasi-distribution function with negative values from experimental data is typically highly cumbersome. Here we propose
a computational approach utilizing a deep generative model, processing three marginals, to construct the bivariate joint quasi-distribution functions. We first apply our model to tackle the
challenging problem of the CHERs, which lacks universal solutions, rendering the problem ground-truth (GT) deficient. To overcome the GT deficiency of the CHER problem, we design optimal synthetic
datasets to train our model. While trained with synthetic data, the physics-informed optimization enables our model to capture the detrimental effect of the thermal fluctuations on nonclassicality, which cannot be obtained from any analytical solutions. This underscores the reliability of our approach. This approach also allows us to predict the Wigner functions subject to thermal noises. Our model
predicts the Wigner functions with a prominent accuracy by processing three marginals of probability distributions. Our approach also provides a significant reduction of the experimental efforts of
constructing the Wigner functions of quantum states, giving rise to an efficient alternative way to realize the quantum state tomography.
\end{abstract}

\maketitle

\onecolumngrid
\section*{Introduction}

Since the birth of quantum theory, the boundary between quantum and classical realms has always been a subject of fascination, spurring extensive research interest
\cite{ballentine_q_statistics_rmp_1970,zurek_q-c_transition_rmp_2003,schlosshauer_q-c_transition_rmp_2005,modi_q-c_boundary_discord_rmp_2012}. One widely adopted approach to distinctly prove genuine
quantumness, as separate from classicality, hinges on the inability of a classical strategy to mimic the statistics of an experiment. The underlying idea of such an approach recognizes that, while a
certain perspective might be intuitively true in the classical realm, it could lead to contradictory outcomes in quantum experiments. As such, when a classical strategy falls short, it strongly
evidences the distinct quantum nature referred to as nonclassicality. This is especially evident in the field of quantum information science. For instance, the renowned example of the experimental
violation \cite{aspect_bell_test_prl_1981,hensen_bell_test_nature_2015,giustina_bell_test_prl_2015,shalm_bell_test_prl_2015} of Bell's inequality \cite{bell_ineq_phys_1964} accentuates the breakdown of
the EPR paradox \cite{epr_paradox_pr_1935} formulated on the tenets of realism and locality. This specific kind of nonclassical correlation resulting in the violation of Bell's inequality is termed Bell
nonlocality \cite{brunner_bell_nonlocal_rmp_2014}.

In line with this idea, the inadequacy of using probability distributions to describe quantum systems indicates the nonclassicality of quantum states. For continuous-variable quantum systems, like nano-mechanical resonators or cavity fields, quantum states are usually described by quasi-distribution functions, such as the Wigner function \cite{wigner_func_pr_1932,
kenfack_wigner_n_cla_jobqsco_2004,ballicchia_wigner_n_cla_appsci_2019} or the Glauber-Sudarshan $P$ representation \cite{glauber_pr_1963,sudarshan_prl_1963,miranowicz_n_cla_pra_2010}. These can be
interpreted as the phase-space representation of quantum states. In the presence of strong quantum coherence, the quasi-distribution function exhibits interference patterns, yielding negative values in
phase space. Hence, such a quantum state lacks a classical analog, highlighting its inherent quantum nature.

This idea can also be employed to identify the nonclassicality of quantum dynamical processes \cite{neill_n_cla_trans_prl_2010,rahimikeshari_process_n_cla_prl_2013,edward_nat_commun_2014,
krishna_process_n_cla_pra_2016,smirne_process_n_cla_qst_2019,simon_process_n_cla_prx_2020,alireza_process_n_cla_prl_2022,dariusz_process_n_cla_sr_2023,budini_process_n_cla_pra_2023} through the newly
emerging canonical Hamiltonian ensemble representation (CHER) \cite{hongbin_process_n_cla_nc_2019,hongbin_cher_sr_2021,muche_process_n_cla_fid_jpcm_2022,yunhua_apaqs_n_cla_fid_prr_2023}. In CHER, the
dynamical behaviors of open quantum systems are described by the ensemble average of unitary evolutions characterized by a Hamiltonian ensemble (HE) \cite{kropf2016effective,
hongbin_process_n_cla_prl_2018,hongbin_disordered_sr_2022}, which comprises a collection of parameterized system Hamiltonian operators associated with a quasi-distribution function. Crucially, the
presence of negative values in the quasi-distribution function indicates the establishment of system-environment quantum correlations, thereby emphasizing the quantum nature of the dynamical process
\cite{hongbin_process_n_cla_nc_2019,hongbin_process_n_cla_prl_2018}. Consequently, the quasi-distribution representation emerges as a cornerstone in characterizing the inherent nonclassicality in
quantum dynamical processes.

Despite the capability in characterizing the nonclassicality, constructing a multivariate joint quasi-distribution function with negative values from experimental data remains a cumbersome task. For
instance, due to the complexity in quantum process tomography, as well as the underlying Lie-algebraic structure of CHER, only specific marginals can be efficiently solved, suggesting that the CHER of
dynamical processes of complex systems can only be built from several marginal distributions. This is reminiscent of a very challenging mathematical issue that has been extensively studied \cite{joint_from_marg_collection_2002,joint_from_marg_jsda_2014,joint_from_marg_2010,joint_from_marg_2019,joint_from_marg_2021}. Particularly, in the context of CHER, complications arise as the quasi-distribution may exhibit negative values, rendering previously developed mathematical techniques unreliable.

Additionally, to construct the Wigner function, one might first determine the complete quantum state through methods such as tomographically measuring rotated quadratures
\cite{smithey_tomography_wigner_prl_1993,breitenbach_tomography_wigner_nature_1997,lvovsky_tomography_wigner_prl_2001}, transformations from different quasi-distribution representations
\cite{deleglise_tomography_nature_2008}, or characteristic-function tomography \cite{fluehmann_c_func_wigner_prl_2020}. Afterward, the Wigner function can be constructed by post-processing the measured
data. Alternatively, one can also directly scan the Wigner function point-by-point in phase space
\cite{lutterbach_measure_wigner_prl_1997,banaszek_measure_wigner_pra_1999,bertet_measure_wigner_prl_2002,brian_measure_wigner_science_2013,falk_measure_wigner_jpb_2022}. All these approaches demand
extensive experimental efforts. Even in the machine-learning-assisted method \cite{ahmed_construct_wigner_prl_2021,ahmed_construct_wigner_prr_2021}, thousands of data points are still required.
Therefore, a reliable novel technique to construct quasi-distribution functions, allowing for negative values, from given marginal distributions is highly desirable. Such a technique would not only be
useful for constructing CHERs, but might also significantly cut down the experimental efforts needed for constructing the Wigner functions of quantum states.

In light of the wide-spreading applications of machine learning techniques in quantum physics, e.g., quantum dynamics~\cite{luchnikov_mach_lear_q_dyna_prl_2020,fanchini_mach_lear_q_dyna_pra_2021,
goswami_mach_lear_q_dyna_pra_2021}, quantum computing~\cite{wise_mach_lear_error_miti_prxq_2021,strikis_mach_lear_error_miti_prxq_2021}, quantum state tomography~\cite{giacomo_mach_lear_qst_np_2018,
sharir_mach_lear_qst_prl_2020,yihui_mach_lear_qst_npjqi_2021,tobias_mach_lear_qst_npjqi_2022}, quantum communication~\cite{wallnofer_mach_lear_appl_prxq_2020,Chin_mach_lear_appl_npjqi_2021}, and
verification of quantum correlations~\cite{lu_mach_lear_sep_ent_pra_2018,ma_mach_lear_bell_ineq_npjqi_2018,canabarro_mach_lear_nonlocality_prl_2019,ren_mach_lear_steerability_pra_2019,
krivachy_mach_lear_nonlocality_npjqi_2020,girardin_mach_lear_sep_appr_prr_2022,hongming_mach_lear_stee_ms_cp_2024} (see also the recent reviews~\cite{carleo_mach_lear_appl_rmp_2019,
karniadakis_mach_lear_appl_nrp_2021,krenn_mach_lear_appl_pra_2023}), here we propose to harness the power of deep generative model (DGM) to generate the corresponding joint quasi-distribution by
processing a series of marginals (Fig.~\ref{fig_illustration_prob}). DGM excels in discerning hidden patterns in expansive datasets, providing invaluable insights. Our DGM integrates the ResNet
structure \cite{kaiming_resnet_2016} to seamlessly generate the bivariate joint quasi-distribution for both problems of CHERs and Wigner functions. To further enhance our DGM's pattern detection within
an image, we have devised a color mapping strategy to encode the joint quasi-distribution into three monochromatic images. To validate our approach and to offer an intuitive result presentation, we
present examples including the CHER of qubit-pair pure dephasing at various temperatures and the Wigner function of a quantum harmonic oscillator exposed to a thermal bath.

On the one hand, we have addressed the issue of lacking ground truth (GT) for predicting the CHERs. By leaning on prior knowledge about the target marginals, synthetic training datasets can be
effectively optimized. On the other hand, our approach offers a streamlined computational alternative, lessening the rigorous experimental demands of Wigner function measurements. Once the well-trained
DGM receives three marginals, which correspond to three probability distributions in real or momentum space, it can generate the desired Wigner function. Additionally, we demonstrate that our
physics-informed approach adeptly constructs high-precision Wigner functions and identifies the detrimental effect of thermal fluctuation on the nonclassicality, especially as temperature rises in the
qubit-pair pure dephasing dynamics. Therefore, our approach provides an experimentally efficient, and reliable, alternative way to realize the conventional quantum state tomography.


\section{Constructing joint quasi-distribution functions from marginals}

\begin{figure}[!ht]
\centering
\includegraphics[width=\columnwidth]{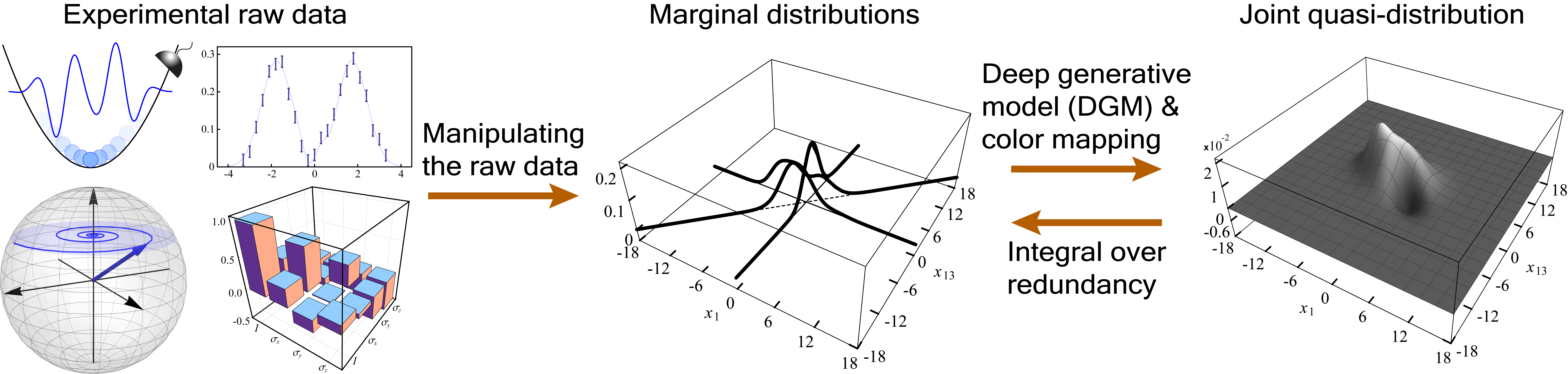}
\caption{\textbf{Construction of bivariate joint quasi-distribution from a set of marginals.}
While the joint quasi-distribution function offers invaluable insights into the nonclassicality, its construction generically requires substantial efforts. In contrast, constructing marginals is
relatively straightforward. Here we develop a DGM integrating color mapping. This DGM enables the construction of bivariate joint quasi-distribution functions by processing three
marginals. These marginals can originate from experimental raw data or be theoretically derived from a prototypical physical model.}
\label{fig_illustration_prob}
\end{figure}

The quasi-distribution representation is a prevalent approach in quantum physics, and its negative values convey a visualized insight into genuine quantum characteristics. However, in many practical
scenarios, its construction can be challenging. For instance, limited by the complexity in quantum process tomography and the underlying Lie-algebraic structure, the construction of the CHERs for
higher-dimensional dynamical processes is typically formidable; moreover, the direct point-by-point measurement of Wigner functions in phase space is experimentally intensive. In contrast, the marginals
derived from the joint quasi-distribution functions are generically standard probability distributions, regardless of the negative values in the quasi-distribution functions. This renders the marginals
physically meaningful and, most importantly, makes them accessible for construction from experimental raw data.

As illustrated in Fig.~\ref{fig_illustration_prob}, given a joint quasi-distribution function $\wp(x_1,x_{13})$, one can, in principle, derive the marginals along any axes by integrating over the
redundant variables, e.g., $\wp(x_1)=\int\wp(x_1,x_{13})dx_{13}$; and the oblique marginal along the diagonal axis can be derived by $\wp(u)=\int\wp(u,v)dv$ along with the change of variables
\begin{equation}
\left\{\begin{array}{l}
u=(x_1+x_{13})/\sqrt{2}\\
v=(-x_1+x_{13})/\sqrt{2}
\end{array}\right..
\end{equation}
On the contrary, although the marginals are experimentally accessible, to construct the desired joint quasi-distribution function $\wp(x_1,x_{13})$ for characterizing the nonclassicality from
three marginals $\wp(x_1)$, $\wp(x_{13})$, and $\wp(u)$ is a long-lasting challenge \cite{joint_from_marg_collection_2002,joint_from_marg_jsda_2014,joint_from_marg_2010,
joint_from_marg_2019,joint_from_marg_2021}.

Motivated by these observations, we are spurred to devise an approach for constructing the desired bivariate joint quasi-distribution functions, which allow for negative values, using three marginals.
These marginals can originate from experimental raw data or be theoretically derived from a prototypical physical model.
We harness artificial intelligence techniques, specifically designing a deep generative model (DGM) integrated with color mapping. Once we have the marginals, whether sourced from experimental raw
data or theoretical calculations, our well-trained DGM efficiently constructs the corresponding joint quasi-distribution function with notable precision.
As concrete paradigms showcasing the capability of the DGM, we apply our approach to resolve the CHER of qubit-pair pure dephasing at various temperatures, and the Wigner function of a quantum harmonic
oscillator exposed to a thermal bath.

\section{Dynamical process nonclassicality characterized by the CHER}

We begin with briefly recalling the theory of dynamical process nonclassicality characterized by the canonical Hamiltonian ensemble representation (CHER). Further details and mathematical supplement can
be found in Supplementary Notes~1 and 2. In this theory, the mathematical toolbox of Hamiltonian ensemble (HE) \cite{kropf2016effective,hongbin_process_n_cla_prl_2018,hongbin_disordered_sr_2022}
$\{(p_{\lambda},\widehat{H}_{\lambda})\}_{\lambda}$ is leveraged to formulate the classical strategy simulating an incoherent dynamics. A HE is a collection of traceless Hermitian operators
$\widehat{H}_{\lambda}\in\mathfrak{su}(n)$ associated with a probability $p_{\lambda}$ of occurrence. It has been shown that if a system can merely establish classical correlations with its environment
during their interaction, then the resulting reduced dynamics $\mathcal{E}_t$ of the system admits a HE decomposition \cite{hongbin_process_n_cla_prl_2018} written as
\begin{equation}
\mathcal{E}_t\{\rho(0)\}=\int p_{\lambda}\widehat{U}_{\lambda}(t)\rho(0)\widehat{U}^{\dag}_{\lambda}(t)d\lambda
\label{eq_ensembe_averaged_dynamics}
\end{equation}
for a certain HE $\{(p_{\lambda},\widehat{H}_{\lambda})\}_{\lambda}$, $\widehat{U}_{\lambda}(t)=\exp(-i\widehat{H}_{\lambda}t)$, and the initial state $\rho(0)$ of the system. Therefore, irrespective of
the inaccessible environmental degrees of freedom in the vast majority of authentic experimental setups, we classify an incoherent dynamics $\mathcal{E}_t$ as classical if it admits a HE-simulation as
Eq.~(\ref{eq_ensembe_averaged_dynamics}). An intuition of this criterion can be acquired from the observation that Eq.~(\ref{eq_ensembe_averaged_dynamics}) can be explained classically as a statistical
mixture of random unitary channels.

On the contrary, $\mathcal{E}_t$ is nonclassical if it admits no HE-simulations. Such nonclassical dynamics arises from the consumption of nonclassical correlations, not reproducible by classical
resources. The nonexistence of simulating HEs can be proven by the necessity to resort to a nonclassical HE encapsulating a quasi-distribution $\wp_\lambda$ with negative values. Consequently, the
negative values in $\wp_\lambda$ is an indicator of the nonclassicality of $\mathcal{E}_t$.

To further promote $\wp_\lambda$ to a characteristic representation over the frequency domain for $\mathcal{E}_t$, referred to as CHER, the ensemble-averaged
simulation~(\ref{eq_ensembe_averaged_dynamics}) has been recast into the Fourier-transform-on-group (FToG) formalism \cite{hongbin_process_n_cla_nc_2019,hongbin_cher_sr_2021}
\begin{equation}
\mathcal{E}^{(\widetilde{L})}_t=\int_\mathcal{G} \wp_\lambda e^{-i\lambda\widetilde{L}t} d\lambda,
\label{eq_ft_on_group}
\end{equation}
where $\mathcal{E}^{(\widetilde{L})}_t$ is the linear map of $\mathcal{E}_t$ expressed with respect to the adjoint representation $\{\widetilde{L}\}$ of the generators of $\mathfrak{u}(n)$.
Noteworthily, this formalism is derived according to the Lie-algebraic structure underlying the HE, rather than the conventional Fourier transform. The FToG~(\ref{eq_ft_on_group}) associates the
quasi-distribution $\wp_\lambda$ with the dynamical process $\mathcal{E}_t$, i.e., $\wp_\lambda\mapsto\mathcal{E}_t^{(\widetilde{L})}$, highlighting the role of $\wp_\lambda$ as a quasi-distribution
representation for $\mathcal{E}_t$. More specifically, to explore the HE-simulation for $\mathcal{E}_t$ means to construct the CHER $\wp_\lambda$ of a joint quasi-distribution by solving the
FToG~(\ref{eq_ft_on_group}).

For a given pure dephasing dynamics $\mathcal{E}_t$ of any dimension, namely the action of $\mathcal{E}_t$ on an arbitrary initial state $\rho_0$ can be explicitly expressed with dephasing factors
$\phi_m(t)$ as
\begin{equation}
\mathcal{E}_t\{\rho_0\}=\left[
\begin{array}{cccc}
\rho_{11} & \rho_{12}\phi_1(t) & \rho_{13}\phi_4(t) & \cdots \\
\rho_{21}\phi_2(t) & \rho_{22} & \rho_{23}\phi_6(t) & \cdots \\
\rho_{31}\phi_5(t) & \rho_{32}\phi_7(t) & \rho_{33} & \cdots \\
\vdots             & \vdots    & \vdots & \ddots
\end{array}
\right],
\label{eq_pure_deph_dyna}
\end{equation}
then the FToG~(\ref{eq_ft_on_group}) can be decomposed into a set of equations governing the marginals along the positive root vectors of
the Lie algebra:
\begin{equation}
\phi_m(t)=\int_{\mathbb{R}^{n-1}}\wp(\mathbf{\lambda})e^{-i(\mathbf{\alpha}_m\cdot\mathbf{\lambda})t}d^{n-1}\mathbf{\lambda},~\mathrm{for~all~positive~root~vectors}~\mathbf{\alpha}_m.
\label{eq_ftog_pure_deph}
\end{equation}
Then the corresponding CHER $\wp(\mathbf{\lambda})$ can only be built from the marginals solved from Eq.~(\ref{eq_ftog_pure_deph}), rendering the direct construction of the CHER intractable. Further
details are shown in Supplementary Note~2.

\section{Quantum pure dephasing dynamics}

To explicitly exemplify the CHER theory, here we specifically focus on pure dephasing dynamics, where only the off-diagonal elements of the density matrix decay over time, while the diagonal
elements remain intact. Although appearing simplistic, the pure dephasing dynamics has proven to be adequate for describing many physical qubit systems on a proper timescale
\cite{nakamura_sc_qubit_prl_2002,petta_st0_qubit_science_2005,foletti_st0_qubit_nat_phys_2009,veldhorst_qubit_control_nn_2014,xianjing_qubit_control_np_2024,childress_nv_center_science_2006,
maze_nv_center_fid_njp_2012,liu_simul_deph_np_2011,liu_simul_deph_nc_2018}. Since the complexity of the CHER problem rises rapidly with the dimension of the Hilbert space of interest, meanwhile the case
of single qubit has been discussed extensively \cite{hongbin_process_n_cla_prl_2018,hongbin_process_n_cla_nc_2019,hongbin_cher_sr_2021,muche_process_n_cla_fid_jpcm_2022,yunhua_apaqs_n_cla_fid_prr_2023},
to design a moderate paradigm striking a balance between feasibility and complexity becomes challenging. Additionally, since the negativity in the CHER stems from the quantum correlations between
the system and its environments, a dynamics model where these correlations give rise to certain exotic phenomena will be crucial for our demonstration.

In view of the above discussions, we consider an extended spin-boson model consisting of a non-interacting qubit pair coupled to a common boson bath with total Hamiltonian
\begin{equation}
\widehat{H}_\mathrm{T}=\sum_{j=1,2}\omega_j\hat{\sigma}_{z,j}/2+\sum_{\mathbf{k}}\omega_{\mathbf{k}}\hat{b}_{\mathbf{k}}^\dagger\hat{b}_{\mathbf{k}}
+\sum_{j,\mathbf{k}}\hat{\sigma}_{z,j}\otimes(g_{\mathbf{k}}\hat{b}_{\mathbf{k}}^\dagger+g_{\mathbf{k}}^\ast\hat{b}_{\mathbf{k}}),
\label{eq_ext_sb_model}
\end{equation}
where $\omega_j$ and $\hat{\sigma}_{z,j}$ are the energy-level spacing and the Pauli $z$ operator of $j$th qubit, respectively, $\omega_{\mathbf{k}}$ and $\hat{b}_{\mathbf{k}}$ are the energy and the
lowering operator of $\mathbf{k}$th environmental mode, respectively, and $g_{\mathbf{k}}$ is the coupling coefficient to $\mathbf{k}$th mode. Note that, even if the two qubits do not interact directly
with each other, they can still establish entanglement via the common boson bath.
This bath-mediated entanglement generation has been discussed using a similar model \cite{royer_quantum_2017,souza_pra_2019}, in which the bath plays a crucial role.

The dynamics is typically solved by transforming the problem into the interaction picture \cite{breuer_textbook,hongbin_3sbm_scirep_2015} with respect to the free Hamiltonian, i.e., the first two terms
in Eq.~(\ref{eq_ext_sb_model}). Then the total system evolves according to the unitary evolution operator
\begin{equation}
\widehat{U}^\mathrm{I}(t)=\mathcal{T}\left\{\exp\left[-i\int_0^t\sum_{\mathbf{k}}\widehat{Z}_{\mathbf{k}}\hat{b}_{\mathbf{k}}^\dagger(\tau)
+\widehat{Z}_{\mathbf{k}}^\dagger \hat{b}_{\mathbf{k}}(\tau)d\tau\right]\right\},
\end{equation}
where $\mathcal{T}$ is the time-ordering operator, $\widehat{Z}_{\mathbf{k}}=g_\mathbf{k}\sum_{j=1,2}\hat{\sigma}_{z,j}$, and
$\hat{b}_{\mathbf{k}}(t)=e^{-i\omega_{\mathbf{k}}t}\hat{b}_{\mathbf{k}}$, respectively.
The density matrix of the qubit pair at a latter time is given by
\begin{equation}
\mathcal{E}_t\{\rho^{(1)}(0)\otimes\rho^{(2)}(0)\}=\Tr_\mathrm{B}\widehat{U}^\mathrm{I}(t)[\rho^{(1)}(0)\otimes\rho^{(2)}(0)\otimes\rho_\mathrm{B}]\widehat{U}^{\mathrm{I}\dagger}(t),
\label{eq_qubit_pair_pure_deph}
\end{equation}
where $\rho^{(j)}(0)$ is the initial state of $j$th qubit, $\rho_\mathrm{B}=\prod_{\mathbf{k}}\exp[-\omega_{\mathbf{k}}\hat{b}_{\mathbf{k}}^\dagger\hat{b}_{\mathbf{k}}/k_\mathrm{B}T]/Z_{\mathbf{k}}$ is
the boson bath in thermal equilibrium at temperature $T$, and $Z_{\mathbf{k}}=\Tr\exp[-\omega_{\mathbf{k}}\hat{b}_{\mathbf{k}}^\dagger\hat{b}_{\mathbf{k}}/k_\mathrm{B}T]$ is the partition function with
Boltzmann constant $k_\mathrm{B}$.

By calculating Eq.~(\ref{eq_qubit_pair_pure_deph}) explicitly, the qubit pair undergoes a pure dephasing dynamics with density matrix characterized by four dephasing factors. By solving the FToG in
Eqs.~(\ref{eq_ft_on_group}) and (\ref{eq_ftog_pure_deph}), the corresponding CHER $\wp(x_1,x_6,x_{13})=\wp_{1,13}(x_1,x_{13})\wp_6(x_6)$ is governed the following equations:
\begin{equation}
\left\{\begin{array}{l}
\phi_1(t)=\exp[i\vartheta(t)-\Phi(t)]=\int_\mathbb{R}\wp_1(x_1)\exp[-ix_1t]dx_1 \\
\phi_9(t)=\exp[-4\Phi(t)]=\int_{\mathbb{R}^2}\wp_{1,13}(x_1,x_{13})\exp[-ix_1t]\exp[-ix_{13}t]dx_1dx_{13} \\
\phi_{13}(t)=\exp[-i\vartheta(t)-\Phi(t)]=\int_\mathbb{R}\wp_{13}(x_{13})\exp[-ix_{13}t]dx_{13} \\
\phi_6(t)=1=\int_\mathbb{R}\wp_6(x_6)\exp[-ix_6t]dx_6
\end{array}\right.,
\label{eq_quantum_dynamics_model}
\end{equation}
with
\begin{equation}
\vartheta(t)=4\int_0^\infty[\mathcal{J}(\omega)/\omega^2](\omega t-\sin\omega t)d\omega,
\end{equation}
\begin{equation}
\Phi(t)=4\int_0^\infty[\mathcal{J}(\omega)/\omega^2]\coth(\omega/2k_\mathrm{B}T)(1-\cos\omega t)d\omega,
\end{equation}
and $\mathcal{J}(\omega)=\sum_{\mathbf{k}}|g_{\mathbf{k}}|^2\delta(\omega-\omega_{\mathbf{k}})$ being the spectral density. In the following, we will demonstrate two types of spectral densities, i.e.,
the family of the super-Ohmic spectral densities, $\mathcal{J}_s(\omega)=\eta\omega^s\omega^{1-s}_\mathrm{c}\exp(-\omega/\omega_\mathrm{c})$ with $s>1$, and the Drude-Lorentz spectral densities,
$\mathcal{J}_\mathrm{DL}(\omega)=(2\eta\gamma/\pi)(\omega/\omega^2+\gamma^2)$, at various temperatures. Both of the spectral densities are extensively used in the literature
\cite{breuer_textbook,hongbin_3sbm_scirep_2015,addis_pra_2014,wilner_prb_2015,ishizaki_pnas_2009,zimanyi_ptrsa_2012,chen_pre_2014}. Further details of the derivation are shown in Supplementary Notes~3.

Note that the first three lines in Eq.~(\ref{eq_quantum_dynamics_model}) determine the three marginals along the $x_1$-, $u$-, and $x_{13}$-axis, respectively; while the last line leads to an
independent component $\wp_6(x_6)=\delta(x_6)$. Therefore, the joint quasi-distribution function $\wp_{1,13}(x_1,x_{13})$ can only be built from these three marginals.
This limitation imposed by the underlying Lie-algebraic structure constitutes the primary obstacle, leading to the lack of efficient approach for constructing the CHERs, but also to the ground-truth
(GT) deficiency for training a supervised learning model. In the following, we explain how to tackle the GT deficiency with synthetic data for training the DGM.

\section{Generation of synthetic training data}\label{sec_data_generation}

One of the most challenges for predicting the CHERs stems from their GT deficiency. Specifically, the lack of efficient approach for solving the FToG~(\ref{eq_ft_on_group}) makes the generation of
adequate training data problematic. To address this issue, we opt to generate synthetic training datasets. In these datasets, the joint quasi-distributions are crafted from a combination of one
bivariate Gaussians with a pair of Gaussians with opposite peaks according to
\begin{equation}
\wp(x_1,x_{13})=p(x_1,x_{13})+A p'(x_1,x_{13})-A p''(x_1,x_{13}),
\label{eq_synthetic_gaussian_data}
\end{equation}
where $p(x_1,x_{13})$, $p'(x_1,x_{13})$, and $p''(x_1,x_{13})$ are three randomly generated (conventional) Gaussians with individual statistical parameters, including mean value, standard deviation, and
correlation coefficient, and $A$ is a random amplitude. The pair of opposite Gaussians is used to mimic potential negativity in the CHERs. Note that $\wp(x_1,x_{13})$ is normalized and the marginals can
be derived analytically. This allows us to efficiently generate the synthetic training datasets associating marginals with joint quasi-distributions.

Additionally, the synthetic datasets are informed by physical principles, making them adaptable. By tweaking relevant parameters, they can be optimized for sufficiently reflecting the target marginals
of a specific quantum dynamics model. This optimization aids in enhancing the proficiency and precision of our DGM. Once the synthetic data are
generated, they can be encoded into the raw data following the process illustrated in Fig.~\ref{fig color map and data shape}.

Furthermore, we set our sights on training the DGM to forecast two separate quantum dynamics models. To achieve this, we generate two unique synthetic training datasets, each optimized for specific
quantum dynamics models. This optimization process begins by examining the profile of the marginals solved from the FToG~(\ref{eq_ft_on_group}) for the intended quantum models. Armed with this
knowledge, we can fine-tune the parameters during the synthetic data generation. This ensures that our synthetic marginals align closely with the nuances of the quantum model to be solved.
Finally, the optimized training dataset consists of 30,000 (31,000) samples, including 10,000 ordinary Gaussians with $A=0$ commonly used in both training datasets and 20,000 (21,000) synthetic
Gaussians with finite $A$ for super-Ohmic (Drude-Lorentz) model, respectively. Further details are shown in Supplementary Note~4.

\section{Data representation and color mapping}

\begin{figure}[!ht]
\includegraphics[width=\columnwidth]{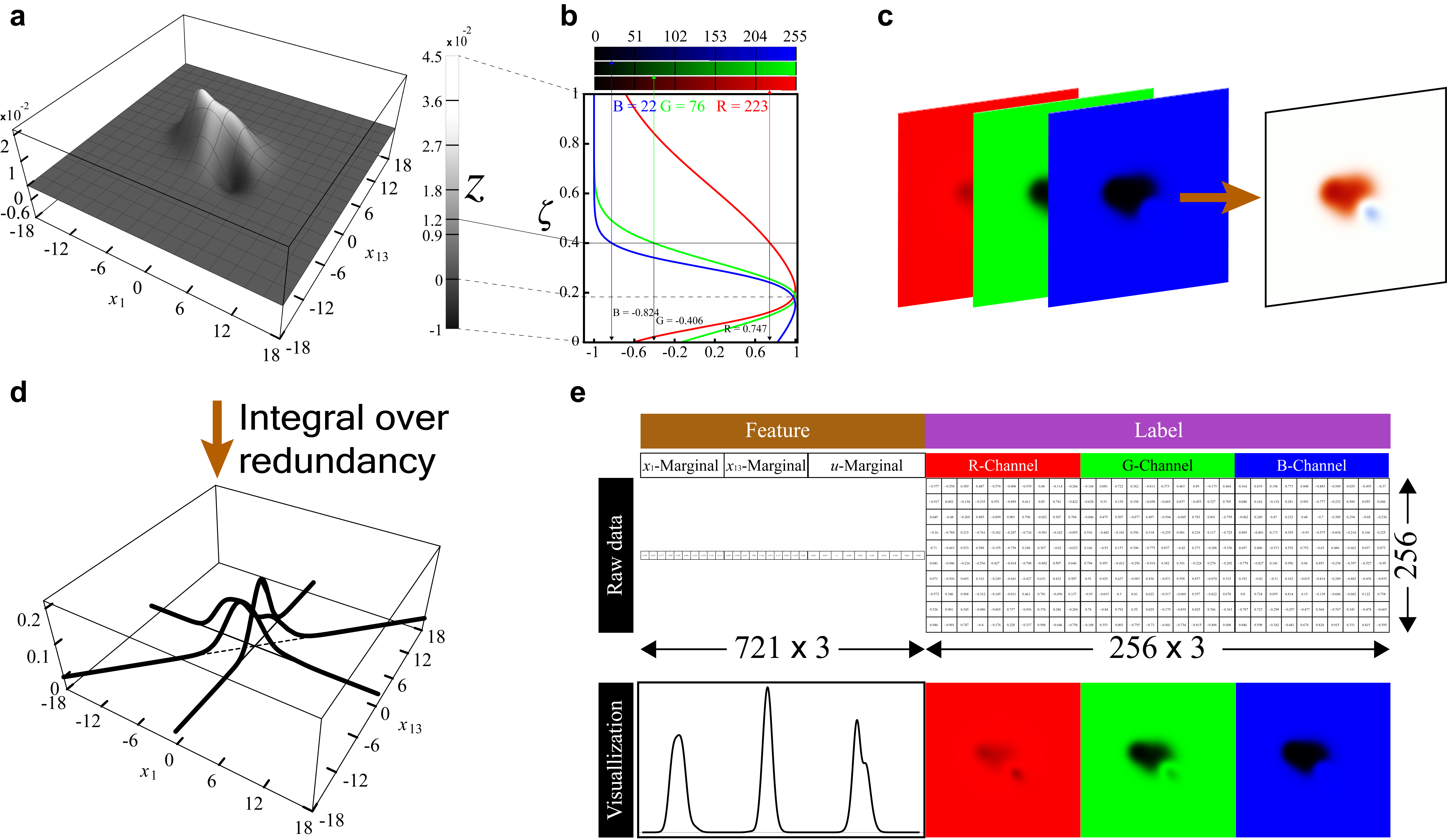}
\caption{\textbf{Generation of training data with a color mapping.} \textbf{a} In the case of the synthetic joint quasi-distribution $\wp(x_1,x_{13})$ in Eq.~(\ref{eq_synthetic_gaussian_data}), the
training dataset can be generated efficiently. \textbf{b}, \textbf{c} The color mapping consists of three functions mapping the (rescaled) height $\zeta$ of the generated training data to a set of RGB
values, leading to three monochromatic images to be learned by the DGM; meanwhile, the monochromatic images can also be merged into a colored image for visualizing the data. \textbf{d} The three
marginals $(\wp(x_1),\wp(x_{13}),\wp(u))$ can be efficiently derived from the synthetic joint quasi-distribution $\wp(x_1,x_{13})$ by integrating over the redundant variables. \textbf{e} Each single
marginal is sampled into 721 pixels and each monochromatic image is sampled into a $256 \times 256$ pixel array. Finally, the three marginals are combined with the three monochromatic images,
constituting the feature and the label of a single training datum, respectively. Here we show the raw data structure fed into our deep generative model and the corresponding visualization.}
\label{fig color map and data shape}
\end{figure}

To train a supervised DGM, it is necessary to prepare a training dataset by associating a label to be learned to a feature to be recognized by the model. The data representation is schematically
illustrated in Fig.~\ref{fig color map and data shape}. The feature of each training datum consists of three marginals $(\wp(x_1),\wp(x_{13}),\wp(u))$, which can be derived efficiently for the
synthetic joint quasi-distribution $\wp(x_1,x_{13})$ in Eq.~(\ref{eq_synthetic_gaussian_data}), as shown in Fig.~\ref{fig color map and data shape}d.

Each marginal is numerically sampled into 721 pixels as the raw data. To take advantage of our DGM in discerning patterns and relationships within the RGB channels of an image, each synthetic joint
quasi-distribution $\wp(x_1,x_{13})$ is translated into three monochromatic images according to a color mapping, as shown in Figs.~\ref{fig color map and data shape}b and c. The color mapping comprises
three functions $(f_\mathrm{R}(\zeta),f_\mathrm{G}(\zeta),f_\mathrm{B}(\zeta))$, mapping the (rescaled) height value $\zeta=(z+0.01)/0.055$ to a set of RGB values, resulting in the three monochromatic images. Note that the color mapping should be optimized according to the training data to be learned (see appendix). Finally, the output monochromatic images serve as the label to be learned by the DGM,
each of which is sampled into a $256 \times 256$ pixel array. The visualization of a single training datum is shown in Fig.~\ref{fig color map and data shape}e. Crucially, the color mapping associates
the positivity and the negativity in the generated data to the R- and B-channels, respectively. This helps our DGM pay particular attention to the negativity in the CHERs, capturing the nonclassical
characteristics. Further details are presented in Supplementary Note~5.

\section{ResNet-based deep generative model}

Our DGM is structured into six stages, each constructed by repeatedly stacking two core building blocks: the identity and deconvolution blocks. As our model aims to produce three monochromatic images
from input marginals, the main stream of both blocks utilizes three deconvolutional layers. These layers extract key patterns from the input marginals and expand them to produce the output monochromatic
images of a quasi-distribution. For the model to handle this intricate task, it requires more depth and additional deconvolutional layers. However, this intensifies the vanishing gradient problem, which
significantly impedes our model's training.

To address this challenge, we have incorporated a shortcut to each block, known as the ResNet structure \cite{kaiming_resnet_2016}. Thanks to the skip connections in the ResNet structure, information
from previous layers is retained, substantially mitigating the vanishing gradient problem. This allows us to deepen our model, enhancing its capabilities. Further insights and comprehensive layout of
our DGM are shown in Supplementary Note~6.

\section{Verifying model performance}

After training the DGMs with the synthetic data, we proceed to verify the performance of the well-trained models on both synthetic datasets. To affirm the reliability of our well-trained DGMs, we have
designed a verification protocol to quantitatively estimate their performance. We compare both the joint quasi-distributions and the derived marginals of the model's prediction with the GT of the test
datum.

During the synthetic data generation procedure explained in Sec.~\ref{sec_data_generation}, we also generate additional 100 testing samples for both DGMs, separate from those in the training datasets.
Once the testing datum is prepared, the three marginals of the testing data are fed into the DGM to output a set of predicted monochromatic images. Then we can compare the predicted monochromatic images
with the GT and estimate the deviation by the pixel-averaged $\mathrm{L}_2$ norm
\begin{equation}
\mathrm{L}_2=\frac{\sqrt{\sum_\mathrm{pixel}|\mathrm{GT}-\mathrm{Prediction}|^2}}{256\times256\times3}.
\end{equation}

Additionally, based on the three predicted monochromatic images, we can reconstruct the corresponding joint quasi-distribution by applying the inverse color mapping, and derive three marginals based on
the predicted joint quasi-distribution. This yields the second quantifier of the performance focusing on the three marginals by estimating the pixel-averaged $\mathrm{L}_1$ norm
\begin{equation}
\mathrm{L}_1=\frac{\sum_\mathrm{pixel}|\mathrm{GT}-\mathrm{Prediction}|}{721}
\end{equation}
for each single marginal. Further details on the verification protocol, numerical results, and average errors are shown in Supplementary Notes~7.

Nevertheless, when forecasting the CHER of a dynamics, the GT-deficiency makes the direct comparison of the joint quasi-distributions infeasible. An alternative verification protocol for the CHERs is
shown in Supplementary Notes~8. Subsequently, we show the predicted CHERs for both quantum spectral densities and estimate the pixel-averaged $\mathrm{L}_1$ norm between marginals derived both from FToG
and the predicted CHERs. This methodology allows us to gauge the reliability of our model in forecasting the CHER, even in the presence of the obstacles of GT-deficiency.

\section{Prediction of CHER}

\begin{figure}[!ht]
\centering
\includegraphics[width=\columnwidth]{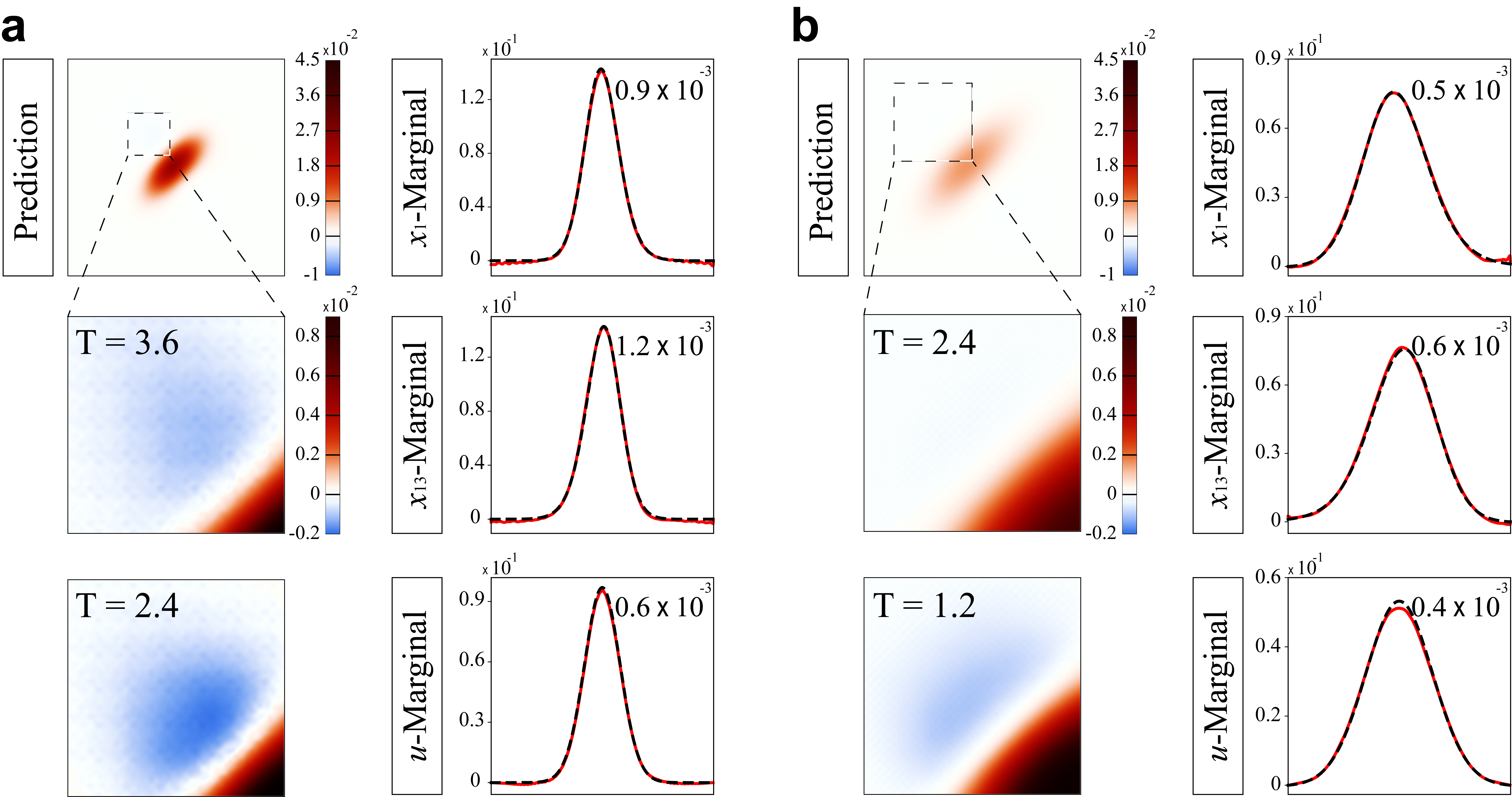}
\caption{\textbf{CHERs predicted by the DGM for both spectral densities.} We show the prediction by our model for the extended spin-boson model~(\ref{eq_ext_sb_model}) with \textbf{a} super-Ohmic spectral
density of Ohmicity $s=2$ and \textbf{b} Drude-Lorentz spectral density. The first panels in the first columns of both \textbf{a} and \textbf{b} show the predicted CHERs. The magnified negative regions are
shown in the following two panels with decreasing temperatures, indicating the nonclassicality of the dynamics. We can also see that our model is capable of recognizing correct physical principle, of which
the negative region strengthens with decreasing temperature, reflecting the detrimental effects of thermal fluctuations. To gauge the reliability of the prediction, we show the results of the verification protocol explained in the previous section in the second columns of both \textbf{a} and \textbf{b}. We shown the comparison between the derived the marginals (red solid curves) from the prediction and the GT marginals (dashed black curves) solved according to Eq.~(\ref{eq_quantum_dynamics_model}). The estimated pixel-averaged $\mathrm{L}_1$ norms are shown at the upper-right corner of each panel.
This underpins the reliability of our model.}
\label{fig_dgm_prediction_dephasing}
\end{figure}

Figure~\ref{fig_dgm_prediction_dephasing} shows the predicted CHERs $\wp_{1,13}(x_1,x_{13})$ for the extended spin-boson model governed by the total Hamiltonian in Eq.~(\ref{eq_ext_sb_model}).
Figure~\ref{fig_dgm_prediction_dephasing}a shows the case of a specific Ohmicity $s=2$ in the super-Ohmic family.
The first panel in the first column of Fig.~\ref{fig_dgm_prediction_dephasing}a shows the prediction for $\wp_{1,13}(x_1,x_{13})$ at temperature $T=3.6$.
We emphasize that this CHER includes a shallow negative region, indicating the nonclassicality of the dynamics.
To clearly exhibit the negative region, we zoom into it and show the magnified CHER in the second panel.
Additionally, we further decrease the temperature to $T=2.4$ and show the enhanced negative region in the last panel.
This is, on the one hand, in line with the intuition that thermal fluctuations are detrimental to the nonclassicality of a quantum system;
on the other hand, this also shows the capability of our model in recognizing correct physical principle, even if the synthetic training datasets do not reflect the effects of thermal fluctuations.

Due to the lack of the GT CHER constructed from Eqs.~(\ref{eq_ft_on_group}) and (\ref{eq_ftog_pure_deph}), to underscore the reliability of the predicted CHER shown in the first column, we have also
numerically derived the marginals (red solid curves) from the predicted CHER and compared them with those of the GT (dashed black curves) solved from the quantum dynamics models in
Eq.~(\ref{eq_quantum_dynamics_model}), as shown in the second column of Fig.~\ref{fig_dgm_prediction_dephasing}a. The pixel-averaged $\mathrm{L}_1$ norm is shown at the upper-right corner of each panel
in the second column. It can be seen that the predictions (red solid curves) fit the GT (dashed black curves) very well.
Furthermore, the errors are two-orders of magnitude lower than the peaking values of the marginals. This affirms the reliability of the predictions.

Figure~\ref{fig_dgm_prediction_dephasing}b shows the prediction for the Drude-Lorentz spectral density. Similarly, we show the predicted CHER $\wp_{1,13}(x_1,x_{13})$ in the first column
of Fig.~\ref{fig_dgm_prediction_dephasing}b, and demonstrate how the negative region strengthens with decreasing temperature.
Noteworthily, the nonclassicality is almost destroyed at a high enough $T=2.4$, revealing a quantum-to-classical transition due to the thermal fluctuations.
We also show the comparison between the derived the marginals (red solid curves) from the prediction and the GT marginals (dashed black curves) solved according to Eq.~(\ref{eq_quantum_dynamics_model})
in the last column of Fig.~\ref{fig_dgm_prediction_dephasing}b with an estimated pixel-averaged $\mathrm{L}_1$ norm shown at the upper-right corner, which underpin the reliability of our model.

\section{Wigner functions and the marginals}

To further demonstrate the potential of our approach, we apply our DGM to another paradigm of Wigner function.
Considering the motional state of a continuous-variable quantum system, it can be equivalently described by either the density matrix $\rho$ or the Wigner function
\begin{equation}
\mathcal{W}(x,p)=\frac{1}{\pi\hbar}\int\bra{x+x^\prime}\rho\ket{x-x^\prime}\exp(-i2px^\prime/\hbar)dx^\prime,
\end{equation}
which is a real-valued quasi-distribution function permitting negative values.
Such a phase-space representation has been proven useful for visualizing quantum states. Particularly, the negative values indicate the inherent quantum characteristics of the states.

Most importantly, a unique property of the Wigner function is that the marginals are standard probability distributions over certain quadratures. For example, given a state vector $\ket{\psi}$ with
wave functions $\psi(x)$ and $\psi(p)$ in real and momentum spaces, respectively, then it can be shown that
\begin{equation}
\mathcal{W}(x)=\int\mathcal{W}(x,p)dp=|\psi(x)|^2
\end{equation}
and
\begin{equation}
\mathcal{W}(p)=\int\mathcal{W}(x,p)dx=|\psi(p)|^2
\end{equation}
are the probability distributions over real and momentum spaces, respectively. Therefore, the marginals, $\mathcal{W}(x)$ and $\mathcal{W}(p)$, are easily experimentally measurable, in contrast to the
point-by-point measurement of the Wigner function $\mathcal{W}(x,p)$ in phase space \cite{lutterbach_measure_wigner_prl_1997,banaszek_measure_wigner_pra_1999,bertet_measure_wigner_prl_2002,
brian_measure_wigner_science_2013,falk_measure_wigner_jpb_2022}.

There have been many approaches developed to construct the Wigner functions from experimental raw data.
For example, the well-developed algorithm of inverse Radon transform can tomographically construct the Wigner function with an extensive measurement over rotated quadratures \cite{smithey_tomography_wigner_prl_1993,breitenbach_tomography_wigner_nature_1997,lvovsky_tomography_wigner_prl_2001}. Other indirect construction of the Wigner function can be transformed from
different quasi-distribution representations \cite{deleglise_tomography_nature_2008}, or characteristic-function tomography \cite{fluehmann_c_func_wigner_prl_2020}.
In addition to these indirect approaches, direct measurement of the Wigner function by the point-by-point scanning in phase space has been experimentally implemented \cite{lutterbach_measure_wigner_prl_1997,banaszek_measure_wigner_pra_1999,bertet_measure_wigner_prl_2002,brian_measure_wigner_science_2013,falk_measure_wigner_jpb_2022}.
All these approaches demand extensive experimental efforts. Therefore, an efficient technique to construct the Wigner function, allowing for negative values, from spare experimental raw data is highly
desirable.

Inspired by the above observation on the relationship between the Wigner function and the corresponding marginals, here we demonstrate the capability of our DGM in constructing the Wigner functions from
experimentally measurable marginals. We specifically consider a quantum harmonic oscillator in coherent or cat sates exposed to thermal bath. The two natural marginals, $\mathcal{W}(x)$ and
$\mathcal{W}(p)$, possess clear physical interpretation as the probability distributions in real and momentum spaces. While the third one $\mathcal{W}(u)$ over the oblique variable $u=(x+p)/\sqrt{2}$ is
accessible in real space after a $\pi/4$-rotation of the quantum states. Further details are shown in Supplementary Note~9.

\section{Generation of datasets and prediction of Wigner functions by the DGM}

\begin{figure}[!ht]
\centering
\includegraphics[width=\columnwidth]{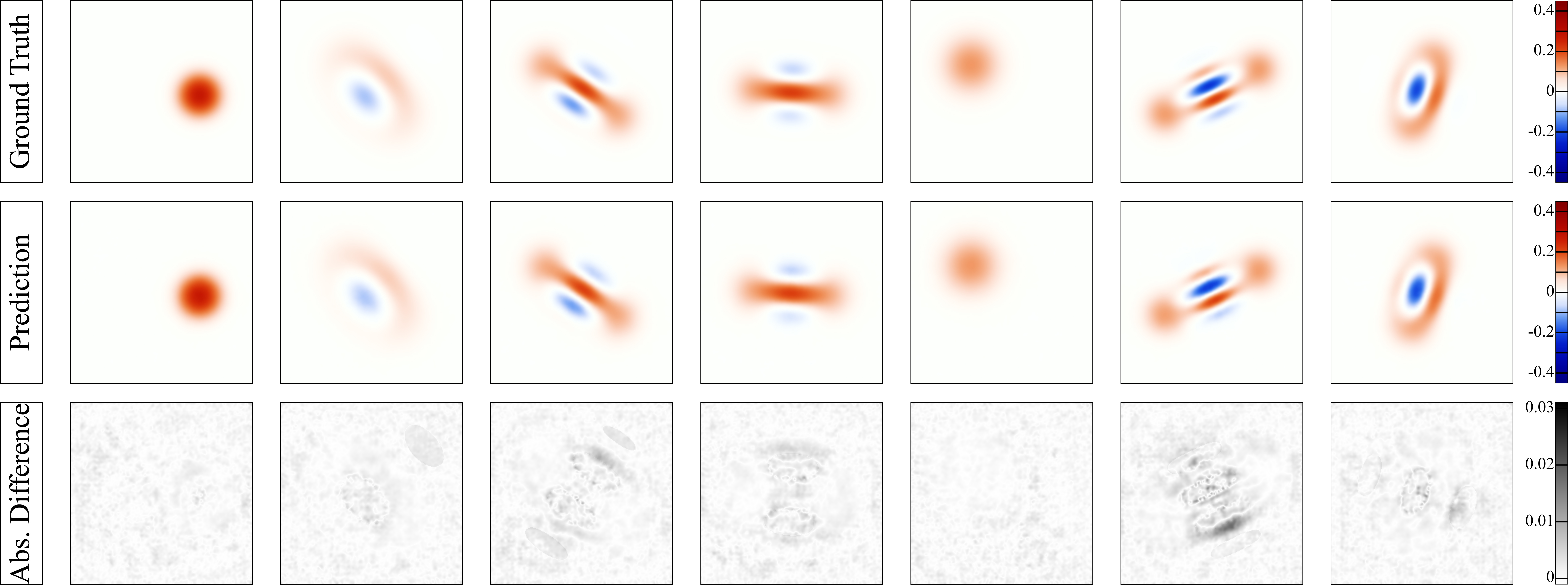}
\caption{\textbf{Verifying the DGM's performance on the Wigner functions.} The top row shows several GT Wigner functions sampled from the testing data used to derive the marginals for the well-trained
DGM. The middle row shows the corresponding predictions. A close comparison reveals the DGM's impressive accuracy, with differences being almost indistinguishable. The bottom row depicts the pixel-wise
absolute differences in grayscale. The DGM precisely predicts noisy coherent states with notable precision. Prominent numerical discrepancies only emerge in regions showcasing significant interference
patterns in the noisy cat states.}
\label{fig_dgm_prediction_wigner}
\end{figure}

As a concrete paradigm, here we consider two types of parameterized states, i.e., the noisy coherent states and the noisy cat states. The thermal noise model \cite{hengna_n_markovianity_scirep_2015},
characterized by $\nu=(1-\mu^2)\bar{n}$, describes the detrimental effects caused by a thermal bath with $\bar{n}$ being the averaged excitation (see appendix and Supplementary Notes~9). Unlike the
situation with the CHER, the Wigner functions of these two parameterized states, as well as the corresponding marginals, can be expressed analytically. Therefore, we can efficiently generate the
necessary datasets for training and testing the DGM. We have generated 16,000 noisy coherent states and 18,000 noisy cat states in the Wigner function training dataset. Further details on how we
generate the datasets and the range of parameters are shown in appendix and Supplementary Notes 9 and 10.

After training our DGM, we apply it to predict the Wigner functions from the three testing data marginals. Figure~\ref{fig_dgm_prediction_wigner} shows several representative samples.
The top row shows the GT Wigner functions used to derive the marginals input into the DGM. The middle row shows the predictions by the DGM.
A close comparison underscores the DGM's accuracy with negligible differences between GT and predictions.
The bottom row highlights the pixel-wise absolute difference between GT and predictions in grayscale.
The DGM's predictions are notably precise for noisy coherent states, as seen in the first and fifth columns.
However, regions with significant interference patterns in the noisy cat states show more pronounced differences.
Further comprehensive quantitative assessment of the model's performance is presented in Supplementary Note~11.

\section{Conclusion and discussion}\label{sec_conclusion}

The quasi-distribution representation permitting negative values is an important approach in characterizing the nonclassical characteristics in a visualized manner.
In this work, we consider two distinct types of nonclassicality, including the nonclassicality of dynamical processes characterized by the CHER,
and the nonclassicality of quantum states characterized by the Wigner function.
The primary difficulties in constructing the multivariate joint quasi-distribution functions lie in the requirement of extensive experimental efforts, as well as the lack of a universal approach.

In this work, we attempt to tackle these difficulties with a more efficient approach using fewer experimental resources.
To this end, we have achieved to leverage the power of deep learning algorithm to establish a DGM for forecasting bivariate joint quasi-distribution functions.
Our DGM incorporates several sophisticated techniques such as the deconvolutional neural networks, the ResNet model, and the color mapping.
Our DGM is efficient in the sense that, it is capable of constructing bivariate joint quasi-distribution functions with a prominent accuracy by processing three marginals of probability distributions,
which are generically experimentally accessible.

The underlying idea of our approach is inspired by an observation of the relationship between the quasi-distribution functions and the corresponding marginals.
First, the desired quasi-distribution functions are experimentally cumbersome.
In contrast, the corresponding marginals are easily accessible experimentally; while the construction of the quasi-distribution functions from given marginals are generically infeasible.
This is reminiscent of a very challenging mathematical issue that has been extensively studied. However, the obstacle is even more stringent in the context of quasi-distribution functions due to the
presence of negative values. Therefore, a computational approach is highly desirable.

To showcase the capability of our approach, we have applied it to resolve both the CHER of qubit-pair pure dephasing at various temperatures, and the Wigner function of a quantum harmonic
oscillator exposed to a thermal bath. The toughest challenge of the former is the lack of universal solution due to the algebraic structure of the FToG governing the CHERs.
This makes the problem a GT-deficient type. To address this challenge, we have designed optimal synthetic datasets based on prior knowledge about the target marginals.
This physics-informed optimization enables our model to capture the detrimental effect of the thermal fluctuations on nonclassicality.
On the other hand, for the latter case of Wigner function, the training datasets can be generated efficiently for both coherent and cat states, even in the presence of thermal noises.
For both cases, our model forecasts the desired bivariate joint quasi-distribution functions with a notable accuracy.
Ultimately, our model provides a streamlined computational approach for constructing the joint quasi-distribution functions.
We believe that this will not only be a critical breakthrough in the analysis of the nonclassicality of complex dynamics, but also considerably lowers the experimental and computational efforts,
giving rise to an efficient and reliable alternative to the conventional quantum state tomography.


\section*{Data availability statement}

The data that support the findings of this study are available upon reasonable request from the corresponding authors.

\section*{Acknowledgments}

The authors acknowledge fruitful discussions with Clemens Gneiting and Franco Nori.
This work is supported by the National Science and Technology Council, Taiwan, with Grants No. MOST 108-2112-M-006-020-MY2, MOST 109-2112-M-006-012, MOST 110-2112-M-006-012, MOST 111-2112-M-006-015-MY3,
MOST 111-2112-M-A49-014, NSTC 112-2112-M-A49-019-MY3, NSTC 112-2123-M-006-001, MOST 112-2314-B-006-011, MOST 110-2224-E-007-003, and MOST 109-2222-E-006-005-MY2,
partially by the Higher Education Sprout Project, Ministry of Education to the Headquarters of University Advancement at NCKU, and partially by the National Center for Theoretical Sciences, Taiwan.

\section*{Author contributions}

H.-B.C. conceived, initiated, and conducted the research.
C.-H.L. developed the methodology with input from B.-Y.T. under the supervision of C.-H.Y. and H.-B.C.
K.-L.L. optimized the model, collected the data, and analysed the results under the supervision of C.-H.Y. and H.-B.C.
P.-Y.L. and Y.-N.C provided the physical meanings of the results.
H.-B.C. drafted the manuscript with input from P.-Y.L. and C.-H.Y.
H.-B.C., Y.-N.C, and C.-H.Y. were responsible for the integration among different research units.

\section*{Conflict of interest}

The authors declare no competing interests.



\section*{Appendix}\label{sec_appendix}



\subsection*{Color mapping}

To deal with the joint quasi-distribution in the sense of an image, we design a color mapping to convert each datum into three monochromatic images. The color mapping consists of three functions
\begin{equation}
\left\{
\begin{array}{l}
f_\mathrm{R}(\zeta) = 2\times\frac{1.148}{(e^{-25(\zeta-(\zeta_0-0.12))}+1)(e^{5(\zeta-(\zeta_0+0.45))}+1)}-1 \\
f_\mathrm{G}(\zeta) = 2\times e^{\frac{-(\zeta-\zeta_0)^{2}}{0.0392}}-1 \\
f_\mathrm{B}(\zeta) = 2\times\frac{1.148}{(e^{25(\zeta-(\zeta_0+0.12))}+1)(e^{-5(\zeta-(\zeta_0-0.45))}+1)}-1
\end{array}
\right.,
\end{equation}
where $\zeta_0$ controls the peaking positions of the RGB curves, and the hight value of the generated data $z\mapsto\zeta$ is rescaled such that $\zeta\in[0,1]$ for most of the training data.
These parameters should be determined according to the statistics of the training data to be learned.
Over-compressing will smear the details in the joint quasi-distributions, and under-compressing will truncate the peaking values of the quasi-distributions.
In the case of Wigner function, we set $\zeta_0=1/2$ and $\zeta=(z+0.45)/0.9$. While for the case of CHER, we set $\zeta_0=1/5.5$ and $\zeta=(z+0.01)/0.055$.
Further advantages of the color mapping on improving our model and the raw data structure are discussed in Supplementary Note~5.

\subsection*{Generation of Wigner function training dataset of noisy states}

In our Wigner function training dataset, we specifically consider two types of parameterized states, i.e., the noisy coherent states and the noisy cat states.
Since the coherent parameter $\alpha\in\mathbb{C}$ is a complex number, its real $\mathrm{Re}[\alpha]$ and imaginary parts $\mathrm{Im}[\alpha]$ are independently sampled in the range $[-2,2]$.
While the relative phase $\theta$ in the noisy cat state is sampled in the range $[0,2\pi)$.
The decoherence effect is characterized by $\mu\in[0.5,1]$, while the thermal noise model \cite{hengna_n_markovianity_scirep_2015} is characterized by $\nu=(1-\mu^2)\bar{n}$ with $\bar{n}\in[0,2]$
denoting the averaged excitation in the thermal bath. We have generated 16,000 noisy coherent states and 18,000 noisy cat states in the Wigner function training dataset.
Since, for both states, the Wigner function $\mathcal{W}(x,p)$ and the three marginals $(\mathcal{W}(x),\mathcal{W}(p),\mathcal{W}(u))$ can be derived analytically, the Wigner function
training dataset can be generated efficiently. The detailed analytic expressions and the range of parameters are presented in Supplementary Notes~9 and 10.


\nocite{apsrev41Control}
\bibliographystyle{apsrev4-1}
\bibliography{deep_learning_nonclassicality}

\end{document}